\documentclass[sigconf]{acmart}

\AtBeginDocument{%
  \providecommand\BibTeX{{%
    \normalfont B\kern-0.5em{\scshape i\kern-0.25em b}\kern-0.8em\TeX}}}

\acmConference[Conference acronym 'XX]{Make sure to enter the correct
  conference title from your rights confirmation emai}{June 03--05,
  2018}{Woodstock, NY}
\usepackage{algorithm}
\usepackage{algorithmic}
\usepackage{subfigure}
\usepackage{makecell}

\usepackage{multirow}
\usepackage{makecell}
\usepackage{subfigure}
\usepackage{multicol}
\usepackage{multirow}
\usepackage{pifont}
\usepackage{xcolor} 
\usepackage{enumerate}
\usepackage{enumitem}

\begin{document}

\title{Revolutionizing Text-to-Image Retrieval as Autoregressive Token-to-Voken Generation}

\author{Yongqi Li$^{1}$, Hongru Cai$^{2}$, Wenjie Wang$^{2}$ Leigang Qu$^{2}$, Yinwei Wei$^{3}$, \\Wenjie Li$^{1}$, Liqiang Nie$^{4}$, Tat-Seng Chua$^{2}$}
\def \authors{Yongqi Li, Hongru Cai, Wenjie Wang, Leigang Qu, Yinwei Wei, Wenjie Li, Liqiang Nie, Tat-Seng Chua}
\affiliation{%
\institution{$^1$The Hong Kong Polytechnic University $^2$National University of Singapore \\ $^3$Monash University $^4$Harbin Institute of Technology (Shenzhen)
\country{}}}
\email{liyongqi0@gmail.com}


\renewcommand{\shortauthors}{Li, et al.}

\begin{abstract}
Text-to-image retrieval is a fundamental task in multimedia processing, aiming to retrieve
semantically relevant cross-modal content. Traditional studies have typically approached this task as a discriminative problem, matching the text and image via the cross-attention mechanism (one-tower framework) or in a common embedding space (two-tower framework). Recently, generative cross-modal retrieval has emerged as a new research line, which assigns images with unique string identifiers and generates the target identifier as the retrieval target. Despite its great potential, existing generative approaches are limited due to the following issues: insufficient visual information in identifiers, misalignment with high-level semantics, and learning gap towards the retrieval target. To address the above issues, we propose an autoregressive voken generation method, named AVG. AVG tokenizes images into vokens, \textit{i.e.}, visual tokens, and innovatively formulates the \textbf{text-to-image retrieval} task as a \textbf{token-to-voken generation} problem. AVG discretizes an image into a sequence of vokens as the identifier of the image, while maintaining the alignment with both the visual information and high-level semantics of the image. Additionally, to bridge the learning gap between generative training and the retrieval target, we incorporate discriminative training to modify the learning direction during token-to-voken training. Extensive experiments demonstrate that AVG achieves superior results in both effectiveness and efficiency. 
\end{abstract}

\ccsdesc[500]{Information systems~Multimedia information systems}
\ccsdesc[500]{Information systems~Multimedia and multimodal retrieval}

\keywords{Image-Text Matching; Cross-Modal Retrieval; Generative Retrieval}



\maketitle

\section{Introduction}
Text-to-image retrieval, as a fundamental task in multimedia processing, has garnered significant attention over the past decade~\cite{wang2017adversarial,zhen2019deep,zheng2020dual}. Its goal is to offer users cross-modal content that goes beyond just text. Despite the fundamental nature of this task, it faces the challenge of bridging the gap between two distinct modalities, text and image, which presents specific research problems.

Current approaches to text-to-image retrieval can be categorized into two groups: 1) The first category, known as the one-tower framework~\cite{chen2020imram,diao2021similarity}, utilizes a cross-attention mechanism to model fine-grained interactions. Recently developed large vision-language models, such as BLIP~\cite{li2022blip}, have shown remarkable capabilities in accurately ranking a small list of images. However, the one-tower architecture falls short in terms of efficiency, making it less suitable for large-scale image retrieval scenarios. 2) The two-tower framework~\cite{chen2021learning,zheng2020dual}, \textit{e.g.}, CLIP~\cite{radford2021learning}, independently maps visual and textual samples into a joint embedding space to calculate cross-modal similarity. This framework trades off some accuracy for increased efficiency, making it excel at retrieving relevant images from extensive image sets. In practice, retrieving a certain number of images from a large-scale set is the initial and essential step to accurately rank a small list of images. Therefore, we concentrate on the fundamental step, to accomplish effective cross-modal retrieval while keeping high efficiency.

\begin{figure}[t]
\setlength{\abovecaptionskip}{0.05cm}
\setlength{\belowcaptionskip}{0cm}
\centering
\includegraphics[width=0.92\linewidth]{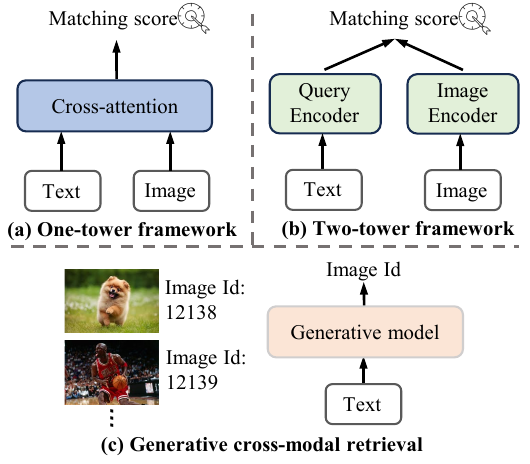}
\caption{Illustrations of three paradigms for cross-modal retrieval. Both the one-tower and two-tower frameworks \textit{match} the text and image for retrieval, while generative cross-modal retrieval \textit{generates} the identifiers of images, \textit{e.g.}, image IDs, as the retrieval results.}
\vspace{-1em}
\label{example}
\end{figure}

\begin{figure*}[t]
\setlength{\abovecaptionskip}{0.05cm}
\setlength{\belowcaptionskip}{-0.3cm}
\centering
  \includegraphics[width=1.0\linewidth]{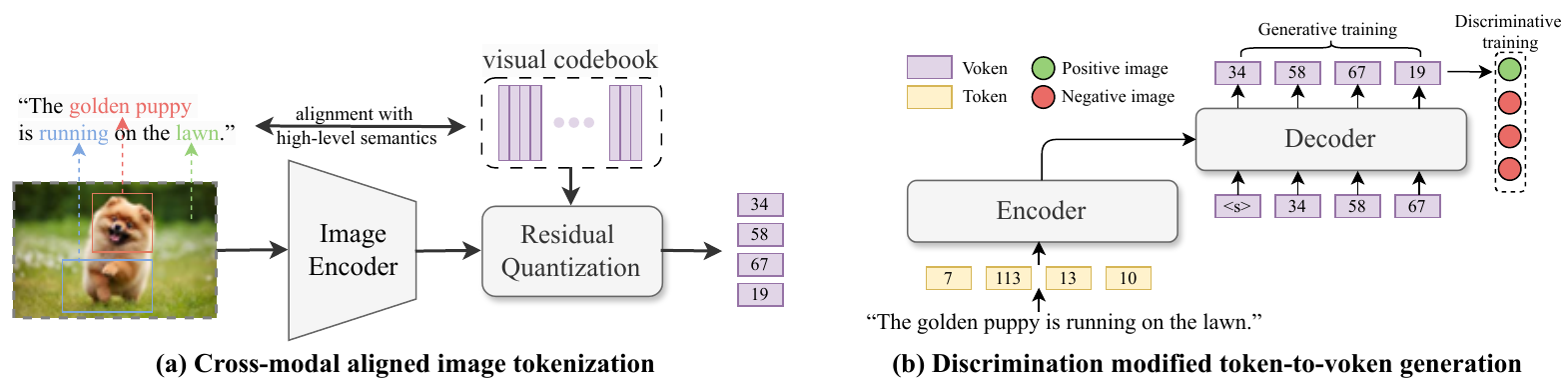}
  \caption{An overview of the proposed AVG method. AVG tokenizes images into a sequence of vokens via the cross-modal aligned image tokenization and devises the discrimination modified token-to-voken generation. }
  \label{method}
\end{figure*}

Generative cross-modal retrieval, a new retrieval paradigm, has recently emerged as a novel research line to cross-modal retrieval~\cite{li2024generative}. As depicted in Figure~\ref{example}, this paradigm initially assigns a unique string identifier, \textit{e.g.}, image Id, to each image and then trains a multimodal large language model~\cite{alayrac2022flamingo}  to generate the corresponding identifier in response to a text query. Unlike traditional frameworks that ``match'' text with images, this approach only receives a textual query as input and ``generates'' the identifiers as the retrieval results during the inference stage. In comparison to the traditional frameworks, the generative paradigm offers several advantages, for example, 1) leveraging powerful generative models, 2) introducing fine-grained interactions in large-scale retrieval scenarios, and 3) maintaining consistent inference efficiency regardless of the size of the image set~\cite{li2024generative}.

Although generative cross-modal retrieval demonstrates great potential, the current approaches are limited in the following aspects: 1)  \textbf{Insufficient visual information in identifiers.} Identifiers play a vital role in generative cross-modal retrieval since the generative model cannot directly analyze the content of an image and instead relies on identifiers as a proxy. However, the existing identifiers~\cite{li2024generative} are predefined and have a weak connection to the visual content of the images. For example, as illustrated in Figure~\ref{example} (c), two distinct images are given similar identifiers, ``12138'' and ``12139'', leading to significant confusion and complexity for the generative model. 2) \textbf{Misalignment with high-level semantics.} Due to the unique nature of cross-modal retrieval, assigning identifiers only based on an image's visual content is inadequate. Much of the visual information is noisy and irrelevant to the textual query, as visual content usually represents low-level information that does not align with high-level semantic meaning. 3) \textbf{Learning gap towards the retrieval target.} Generative learning focuses on predicting the correct image (identifier), while the retrieval task aims to obtain a high-quality ranked list. This creates a learning gap between generative learning and the retrieval target, leading to a deviated optimization direction to the optimal learning direction.

In this study, we propose an autoregressive ``voken'' generation method, named AVG, for text-to-image retrieval. As illustrated in Figure~\ref{method}, AVG tokenizes images into a sequence of ``vokens'' (\textit{i.e.} visual tokens), innovatively reformulating the text-to-image retrieval task into a token-to-voken generation problem. 
1) To tackle the limitations of the absence of visual information and misalignment with high-level semantics, AVG devises a cross-modal aligned image tokenizer. Specifically, a semantic alignment loss is integrated into the standard visual quantization process, regulating the image vokens to align well with both the low-level visual information and high-level semantics simultaneously. 
Compared to the predefined string identifiers~\cite{li2024generative}, the vokens are learnable from images' visual and semantic content, serving superior identifiers for cross-modal retrieval. 
2) Thereafter, we extend the vocabulary of a language model by including the well-learned vokens, and then train the model to accept a textual query and generate the target image's vokens as output. 
To mitigate the learning gap between generative training and the intended retrieval target, we introduce an aided discriminative loss to empower the generative language model with discriminative ranking capabilities. 
The experiments show that AVG largely surpasses previous generative cross-modal retrieval methods with an average improvement of 14.5 \textit{w.r.t.} Recall@10 on Flickr and MS-COCO datasets. 
Compared with the classical two-tower method, CLIP~\cite{radford2021learning}, AVG achieves superior results while maintaining a 4$\times$ efficiency improvement.


The key contributions in this work are as follows:
\begin{itemize}[leftmargin=*]
  \item By tokenizing images into vokens, we innovatively formulate the text-to-image retrieval task into a token-to-voken generation problem, facilitating the utilization of powerful language models in generative cross-modal retrieval. 
  
  \item We propose a cross-modal aligned image tokenizer, injecting both low-level visual information and high-level semantics into vokens as image identifiers for generative cross-modal retrieval. 
  
  \item We identify the learning gap between generative training and the intended cross-modal retrieval objective, and propose to modify the deviated learning direction via aided discriminative training. 
  
  \item Extensive experiments on two benchmarks validate the superiority of AVG on retrieval effectiveness and efficiency. 
  
\end{itemize}


\section{Related Work}
\subsection{Cross-modal Retrieval}
Cross-modal retrieval (text-image matching) ~\cite{yang2023knowledge, dong2023giving,li2023towards,liu2023multi} is a fundamental task in multimedia processing, which can be categorized into one-tower framework and two-tower framework according to the modality interactions. The one-tower framework~\cite{chen2020imram,diao2021similarity,lee2018stacked,qu2021dynamic} incorporates fine-grained cross-modal interactions to achieve matching between fragments, such as objects and words. For example, the pioneering work SCAN~\cite{lee2018stacked} utilized cross-modal interaction to infer latent region-word alignments. Diao et al.~\cite{diao2021similarity} inferred multi-level semantic correlations in a similarity graph and used the attention mechanism to filter out noisy alignments. In the two-tower framework~\cite{chen2021learning,faghri2017vse++,zheng2020dual,qu2020context}, images and texts are independently mapped into a joint feature space, where semantic similarities are calculated using cosine functions or Euclidean distances. For example, Frome et al.~\cite{frome2013devise} proposed a method that represents visual and textual instances in a modality-agnostic space to assess their semantic relevance. To enhance discriminative power, Zheng et al.~\cite{zheng2020dual}  argued that the commonly used ranking loss is ineffective for large-scale multi-modality data, and they introduced a new instance loss that exploits intra-modal data distribution in an end-to-end manner.

It is important to note that the one-tower framework and two-tower framework are not directly comparable. The one-tower framework prioritizes fine-grained interaction at the expense of efficiency, making it less suitable for retrieval with large-scale image sets. Both the one-tower framework and the two-tower framework formulate cross-modal retrieval as a discriminative problem, which relies on discriminative loss and negative samples~\cite{li2023your}. Differently, we formulate the text-to-image retrieval task as the token-to-voken generation problem. Our work is designed to be applicable to large-scale image sets with high efficiency, making it comparable to the two-tower framework.

\subsection{Generative Retrieval}
Generative retrieval is an emerging paradigm in text retrieval that involves generating identifier strings of passages as the retrieval target. Identifiers play an important role in generative retrieval, which could reduce the volume of irrelevant information and facilitate easier memorization and learning for the model~\cite{li2023multiview}. Different types of identifiers have been explored in various search scenarios, including passage titles (Web URLs), numeric IDs, and substrings of passages, as shown in previous studies~\cite{de2020autoregressive, tay2022transformer,bevilacqua2022autoregressive,ren2023tome}. Although the text is naturally discrete, there are studies~\cite{sun2024learning,zeng2023scalable} demonstrating that relearning a codebook to discretize the documents is effective for text generative retrieval. Regrettably, the above identifiers are ineffective for images in the context of cross-modal retrieval, due to the unique challenges of bridging the gap between two distinct modalities in this task.

Beyond the text retrieval, Li et al.~\cite{li2024generative} first proposed the generative cross-modal retrieval paradigm. However, they used predefined strings as identifiers of images, which lacked the visual information of images. IRGen~\cite{zhang2023irgen} applied the generative retrieval into the image-to-image retrieval task and also adopted the image tokenization technique. However, IRGen is ineffective in text-to-image retrieval due to the misalignment with semantics in the image tokenization stage. Besides, they also overlooked the learning gap between generation training and the retrieval target.

\subsection{Image Tokenization}
In 2017, \citeauthor{van2017neural}~\cite{van2017neural} proposed the Vector Quantized Variational Autoencoder (VQ-VAE), a model designed to learn the low-dimensional discrete representation of images and autoregressively generate the images. Building upon this work, VQ-VAE2~\cite{razavi2019generating} incorporated a hierarchy of discrete representations. Subsequently, VQ-GAN~\cite{esser2021taming} further enhanced the perceptual quality of reconstructed images by incorporating adversarial and perceptual loss. Additionally, ViT-VQGAN~\cite{yu2021vector} introduced the powerful transformer backbone to the VQGAN model. Lastly, RQ-VAE~\cite{lee2022autoregressive} utilized Residual Quantization~\cite{juang1982multiple} to iteratively quantize a vector and its residuals, representing the vector as a stack of tokens. It is important to note that the learned tokens are merely a by-product of the image generation process.

The above image tokenization methods mainly serve for autoregressive image generation, and thus only consider the visual content for quantization. They usually discretized an image into a long sequence of vokens, which cannot serve as effective identifiers for generative retrieval. Considering the nature of cross-modal retrieval, we developed the cross-modal aligned image tokenization to consider the semantic information for tokenization.
\section{Method}
Our proposed AVG consists of two stages:

\textbf{Cross-modal Aligned Image Tokenization}: This step aims to develop an effective image tokenizer, which discretizes an image into a sequence of vokens based on not only its visual content but also the high-level semantics.

\textbf{Discrimination Modified Token-to-Voken Generation}: By discretizing images into vokens, the image retrieval task is formulated as the token-to-voken generation problem. To mitigate the gap between generation and retrieval, a discriminative loss is introduced to modify the learning goal of the generative training.
\subsection{Cross-modal Aligned Image Tokenization}
Unlike text, which naturally breaks down into discrete units, images are continuous. To generate an image in an autoregressive manner, the first step is to discretize the image. With the success of autoregressive generation, the concept of image tokenization has been investigated through methods like VQ-VAE~\cite{van2017neural} and RQ-VAE~\cite{lee2022autoregressive}. Essentially, these approaches utilize a variational autoencoder to learn a visual codebook, which is then used to map image patches onto a sequence of vectors corresponding to entries in the visual codebook. 

\begin{figure}[t]
\centering
\includegraphics[width=1.0\linewidth]{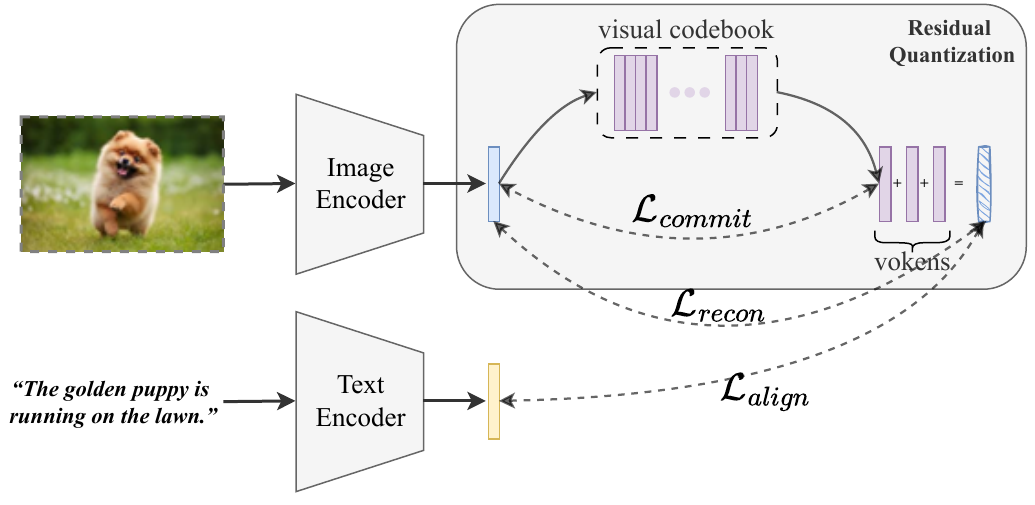}
\vspace{-2em}
\caption{Illustration of the cross-modal aligned image tokenization. We introduce the semantic alignment loss, $\mathcal{L}_{align}$, to guarantee the learned vokens are assigned based on an image's visual information and high-level semantics.}
\label{cross-modal alignment}
\end{figure}

Despite the success of image tokenizers in autoregressive image generation, they are not well-suited for cross-modal retrieval tasks for several practical reasons: 1) The tokenization process results in an excessively long sequence of tokens. For instance, VQ-VAE divides a 128*128 image into 32*32 patches, creating a total of 1,024 tokens. It becomes challenging for the autoregressive model to generate these 1,024 tokens accurately without errors. 2) Image tokenizers are spatially aware that could be harmful to cross-modal retrieval. For example, the initial tokens of an image correspond to its top-left region, which may not be semantically related to the textual query. 3) Current image tokenizers focus primarily on the visual aspects of images and neglect the high-level semantics, which is crucial for effective cross-modal retrieval. 

Therefore, we are motivated to develop the image tokenizer specifically for the cross-modal retrieval task, as shown in Figure~\ref{cross-modal alignment}. Denote an image as $i$ and its associated textual query as $t$. We first input the image and the textual query into a visual encoder and text encoder to obtain the image vector and text vector, respectively, formulated as follows,
 \begin{equation}  \label{eqn1}
  \left\{
   \begin{aligned}
& \textbf{v}_i = \textbf{Encoder}_{v}(i), \\
& \textbf{v}_t = \textbf{Encoder}_{t}(t). 
   \end{aligned}
   \right.
 \end{equation}
$\textbf{Encoder}_{v}$ and $\textbf{Encoder}_{t}$ denote the visual encoder and text encoder, respectively. In practice, we apply the pretrained transformer-based visual encoder and text encoder to warm up the networks.  

\textbf{Residual quantization (RQ)}. We adopt the residual quantization~\cite{lee2022autoregressive} to quantize the image vector $\textbf{v}_i$. We first define a visual codebook $\textbf{C} = \{\textbf{voken}_1, \textbf{voken}_2, ..., \textbf{voken}_N\}$, where $\textbf{C}$ is a randomly initialized matrix and each line indicates a voken vector. Thereafter, we perform $M$ steps to sequentially select $M$ voken vectors from the visual codebook $\textbf{C}$ for representing an image. In the first step, the residual vector is initialized as $\textbf{r}_1 = \textbf{v}_i$. The residual vector is then used to search the nearest voken vector from the visual codebook $\textbf{C}$, denoted as $\textbf{voken}_1^i$.  In the second step, the residual vector is updated as $\textbf{r}_2 = \textbf{r}_1 - \textbf{voken}_1^i$. And then, the current residual vector is used to search the nearest voken vector, denoted as $\textbf{voken}_2^i$. For the quantization step $j \in \{1, 2, ..., M\}$ of the image $i$, the residual quantization is formulated as,
 \begin{equation}  \label{eqn2}
  \left\{
   \begin{aligned}
& \textbf{r}_j = \textbf{r}_{j-1} - \textbf{voken}_{j-1}^i, \\
&  \textbf{voken}_{j}^i = <\textbf{r}_j, \textbf{C}>, 
   \end{aligned}
   \right.
 \end{equation}
 where $<,>$ denotes the search of the nearest voken vector from $\textbf{C}$ and $\textbf{voken}_{j}^i$ is the searched voken vector in the step $j$.

By repeating the above process in Eqn. (\ref{eqn2}) for $M$ times, we could obtain $M$ voken vectors. Given an image $i$, we discretize it as a sequence of $M$ vokens $\{\textbf{voken}_1^i, \textbf{voken}_2^i, ..., \textbf{voken}_{M}^i\}$. We define $\hat{\textbf{z}}^{(d)}=\sum_{j=1}^d\textbf{voken}_j^i, 1 \leq d \leq M, $ as the partial sum of up to $d$ voken embeddings, and $\hat{\textbf{z}}^{(M)}$ as the quantized vector. We then forward the $\hat{\textbf{z}}^{(M)}$ into the Decoder layer (MLP layer) and define the reconstruction loss as,
 \begin{equation}  \label{eqn3}
  \left\{
   \begin{aligned}
& \hat{\textbf{v}}_i =\textbf{MLP}(\hat{\textbf{z}}^{(M)}), \\
&  \mathcal{L}_{recon} = || \textbf{v}_i - \hat{\textbf{v}}_i||_2^2, 
   \end{aligned}
   \right.
 \end{equation}
where $\hat{\textbf{v}_i}$ is the reconstructed image vector. We adopt the shallow MLP as the Decoder layer because our focus is on image retrieval rather than image generation. Therefore, our goal is to reconstruct the image vector rather than the entire image itself. The reconstruction loss guarantees that the discrete vokens keep the original visual information as much as possible. Besides, the commit loss is defined as follows,
 \begin{equation}  \label{eqn4}
   \begin{aligned}
 \mathcal{L}_{commit} = \sum_{d=1}^{M}|| \textbf{v}_i - sg(\hat{\textbf{z}}^{(d)})||_2^2, 
   \end{aligned}
 \end{equation}
where $sg()$ is the stop-gradient operator. The commit loss aims to make $\hat{\textbf{z}}^{(d)}$ sequentially decrease the quantization error of $\textbf{v}_i$ as d increases. 

\textbf{Cross-modal alignment}. The above RQ process only guarantees the learned vokens keep the original visual information as much as possible, but overlooks the essential high-level semantics. To handle this problem, we introduce the semantic alignment loss, defined as,
 \begin{equation}  \label{eqn5}
   \begin{aligned}
 \mathcal{L}_{align} = ||\hat{\textbf{v}}_i -  \textbf{v}_t||_2^2. 
   \end{aligned}
 \end{equation}
The alignment loss injects semantics to the visual codebook by constraining the distance between the reconstructed image vector $\hat{\textbf{v}}_i$ and text vector $\textbf{v}_t$.

The final loss to train the image tokenizer is defined as the sum of the reconstruction loss, the commit loss, and the semantic alignment loss, as follows,
 \begin{equation}  \label{eqn6}
   \begin{aligned}
   &\mathcal{L}_{tokenizer}= &\mathcal{L}_{recon} + \mathcal{L}_{commit} + \mathcal{L}_{align}.
   \end{aligned}
 \end{equation}

\subsection{Discrimination Modified Token-to-Voken Generation}
Since we could discretize an image $i$ into a sequence of vokens and these tokens could well represent the image's visual content and high-level semantics, the task of text-to-image retrieval can naturally be reinterpreted as token-to-voken generation problem.

\textbf{Generative training}. The autoregressive language models, such as T5~\cite{raffel2020exploring} and LLaMA~\cite{touvron2023llama}, denoted as $\textbf{AM}$, only support token-to-token generation. Therefore, we first expand the language models' vocabulary by adding the vokens in the learned visual codebook $\textbf{C}$ into the original token list. It is notable that we also copy the voken's embedding into the vocabulary to take full advantage of the embedded semantics, which will benefit the following generative training. We then continue training the generative language model using the next ``voken'' prediction, formulated as follows,

\begin{equation}  \label{eqn7}
   \begin{aligned}
   &\mathcal{L}_{gen} = 
   &-\sum_{j=1}^{N}\log p_{\theta}({voken_j}|token_{<L};voken_{<j}),
   \end{aligned}
 \end{equation}
where $voken_{<j}$ denotes the vokens from an image $i$, $voken_{<L}$ denotes the tokens after tokenizing the text $t$, and $L$ is the length of the sequence of tokens. $\theta$ is the parameters of the generative language model. After the generative training, the autoregressive language model $\textbf{AM}$ could conduct associations between text (tokens) and image (vokens). 

\textbf{Discriminative training}. As we claimed in the Introduction, there is a notable discrepancy between generative training and the intended retrieval target. To mitigate this learning gap, we introduce a discriminative loss to modify the learning direction of the autoregressive model $\textbf{AM}$. We first prompt the language model to generate several images (vokens) via beam search, as follows,
\begin{equation}  \label{eqn8}
  \left\{
   \begin{aligned}
   &\mathcal{I}_r = \textbf{AM}(token_{<L};b), \\
   &\mathcal{I}_r = \{I_1, I_2, ..., I_{b}\},
   \end{aligned}
      \right.
 \end{equation}
where $\mathcal{I}_r$ denotes the retrieved image list and $b$ is the beam size. The $\textbf{AM}$ receives the tokens and then generates several sequences of vokens, and each sequence of vokens could correspond to one image. For each image $I_j (1 \leq j \leq b)$ in $\mathcal{I}_r$, we could define its score, $s_j$ as the sum of the corresponding vokens' generation probabilities. 

If there is no relevant image in $\mathcal{I}_r$, we will randomly replace one image in  $\mathcal{I}_r$ with the positive image to the textual query. We utilize the typical contrastive loss to optimize over the image list $\mathcal{I}_r$, formulated as,
\begin{equation}  \label{eqn9}
   \begin{aligned}
   &\mathcal{L}_{dis}=-\log\frac{\exp(s_p)}
{\sum_{j=1}^{b}\exp(s_j)},
   \end{aligned}
 \end{equation}
 where $s_p$ denotes the positive image's score. This discriminative loss guides the model to learn a good rank list rather than a single image.

 Finally, we optimize the autoregressive language model $\textbf{AM}$ via the sum of the generation loss and the discrimination loss, as follows,
 \begin{equation}  \label{eqn10}
   \begin{aligned}
   &\mathcal{L}_{AM}= &\mathcal{L}_{gen} + \mathcal{L}_{dis}.
   \end{aligned}
 \end{equation}

\subsection{Inference}
After training the image tokenizer and autoregressive model $\textbf{AM}$, AVG is able to retrieve an image from the image set easily. When given a textual query, AVG inputs it into the autoregressive model $\textbf{AM}$, which then predicts a sequence of vokens that could correspond to images. To obtain a rank list of images, the beam search is applied. To prevent the model from generating invalid vokens that do not correspond to images in the test set, we apply constrained generation technique~\cite{de2020autoregressive}.

\section{Experiments}
We conduct experiments to answer four research questions:

\noindent$\bullet\quad$\textbf{RQ1:} How does our proposed AVG compare in performance to previous generative cross-modal retrieval methods?

\noindent$\bullet\quad$\textbf{RQ2:} How does the AVG compare in effectiveness and efficiency to the classical one-tower and two-tower frameworks?

\noindent$\bullet\quad$\textbf{RQ3:} How do different components contribute to the performance of AVG?

\noindent$\bullet\quad$\textbf{RQ4:} How does AVG perform with different key settings, such as the visual codebook, voken length, and model size?
\subsection{Datasets and Baselines}
\textbf{Datasets.} We evaluated the proposed autoregressive voken generation method, AVG, on two widely-used datasets: Flickr30K~\cite{young2014image} and MS-COCO~\cite{lin2014microsoft}. Flickr30K comprises 31,783 images, each paired with five human-annotated sentences. We utilized the data split employed by ~\citet{li2019visual}, with 29,783 images for training, 1,000 for validation, and 1,000 for testing. MS-COCO consists of 123,287 images, each accompanied by five annotated sentences. We followed the dataset split proposed in ~\cite{lee2018stacked}, using 113,287 images for training, 5,000 for validation, and 5,000 for testing. In line with previous studies~\cite{young2014image, chen2021learning}, we evaluated our method using the standard recall metric $R@K$, with $K$ set to 1, 5, and 10.

\textbf{Baselines.} As generative cross-modal retrieval is a relatively new paradigm, there are only a few methods that serve as baselines. The first method, GRACE~\cite{li2024generative}, is the pioneering generative approach to cross-modal retrieval. It explores various identifier types for images, such as Numeric Id, String Id, Semantic Id, and Structured Id. It is worth noting that GRACE also explores the concept of "Atomic Id," which assigns each image a unique vector in the vocabulary. However, the "Atomic Id" approach is not a generative paradigm, as it necessitates the maintenance of a large embedding matrix for all images. Consequently, we did not include the "Atomic Id" as the generative retrieval baseline. The second method, IRGen~\cite{zhang2023irgen}, is proposed for image-to-image retrieval and also utilizes the image tokenization technique. We adapted IRGen for the text-to-image task by substituting the image input with text input. Besides, we also compared AVG with the other classical methods, BLIP~\cite{li2022blip} and CLIP~\cite{radford2021learning}. We analyzed the effectiveness and efficiency of three paradigms in Section 4.4.
\subsection{Implement Details}
We train the model on 4$\times$24GB NVIDIA A5000 GPUs. For the training of image tokenization, we set a learning rate of 0.0001 and a batch size of 2048 for each GPU, while the learning rate was 0.001 and the batch size was 128 for the generative training.  For the image tokenizer, we evaluated the size of the visual codebook, $N$ in \{256, 512, 1024, 2048\}, and the voken length $M$ in \{2, 4, 6, 8, 10\}. For the autoregressive model,  we evaluated T5-base, T5-large, T5-XL, and LLaMA-7b. In the inference stage, we looped the beam size in \{1, 5, 10, 20, 30, 40, 50\} for beam search.

\begin{table}[t]
\renewcommand\arraystretch{1}
  \centering
 \caption{Performance comparison between our proposed AVG and other generative cross-modal retrieval methods. The best results in each
group are marked in Bold, and * implies the improvements over the second-best baseline are statistically significant (p-value < 0.01) under one-sample t-tests.}
 \vspace{-0.5em}
\resizebox{0.8\linewidth}{!}{
    \begin{tabular}{cccc}
    \toprule
    \multicolumn{1}{c}{\multirow{1}*{Method}}
         &R@1&R@5&R@10\cr\toprule
         \multicolumn{4}{c}{Flickr30K}\cr
    \toprule
    GRACE~\cite{li2024generative} (Numeric Id)&22.5&28.9&29.4 \cr
    GRACE~\cite{li2024generative} (String Id)&30.5&39.0&40.4 \cr
    GRACE~\cite{li2024generative} (Semantic Id)&22.9&34.9&37.4 \cr
    GRACE~\cite{li2024generative} (Structured Id)&37.4&59.5&66.2\cr
    IRGen~\cite{zhang2023irgen}&49.0&68.9&72.5 \cr
    AVG (Ours)&$\textbf{62.8}^*$&$\textbf{85.4}^*$&$\textbf{91.2}^*$ \cr
    \toprule
         \multicolumn{4}{c}{MS-COCO (5k)}\cr
    \toprule
    GRACE~\cite{li2024generative} (Numeric Id)&0.03&0.14&0.28 \cr
    GRACE~\cite{li2024generative} (String Id)&0.12&0.37&0.88 \cr
    GRACE~\cite{li2024generative} (Semantic Id)&13.3&30.4&35.9 \cr
    GRACE~\cite{li2024generative} (Structured Id)&16.7&39.2&50.3\cr
    IRGen~\cite{zhang2023irgen}&29.6&50.7&56.3 \cr
    AVG (Ours)&$\textbf{31.3}^*$&$\textbf{58.0}^*$&$\textbf{66.5}^*$ \cr\toprule
    \end{tabular}}
    \vspace{-1em}
    \label{tab:overall performance}
\end{table}

\subsection{Overall Performance (RQ1)}
To answer the question RQ1, we conducted experiments on Flickr30K and MS-COCO (5k) to compare AVG with previous generative cross-modal retrieval methods. The results are presented in Table~\ref{tab:overall performance},
from which we have the following observations.

\noindent$\bullet\quad$ Among the three generative cross-modal retrieval approaches, GRACE performs poorly compared to IRGen and AVG. This is because GRACE determines an image's content without taking its visual content into consideration. This is evident from the varying performance of different identifier types of GRACE. It has been observed that the Numeric Id and String Id perform poorly compared to the Semantic Id and Structured Id, as they are only partially related to the image's content. However, even the Semantic Id and Structured Id still underperform IRGen and AVG, as they are predefined, whereas IRGen and AVG learn image tokenizers to automatically assign identifiers based on the visual content.

\noindent$\bullet\quad$ Despite both learning the image tokenizer for identifiers, IRGen still underperforms AVG for two main reasons. Firstly, during the training of the image tokenizer, IRGen only considers an image's visual content as input, and residual quantizes the visual embedding vector. This approach is not suitable for the cross-modal retrieval task, as high-level semantic information is crucial. In contrast, AVG introduces semantic alignment during the training of the image tokenizer, aligning tokens and vokens in advance, which can benefit the subsequent token-to-voken generation. Secondly, AVG incorporates a discriminative loss in the generative training phase, which helps to bridge the gap between generative training and the retrieval target, resulting in improved image ranking results.

\noindent$\bullet\quad$  Comparing the results from the two datasets, we discovered that all methods performed better on Flickr than on MS-COCO. The primary reason for this discrepancy is the difference in the sizes of the image sets. On Flickr, the model is tasked with retrieving relevant images from a pool of 1K images, whereas the size of the pool on MS-COCO is 5K. The larger image sets present a greater challenge for the models. Additionally, it was observed that GRACE (Numeric Id) and GRACE (String Id) did not yield successful results on MS-COCO. These methods rely on the model memorizing the associations between identifiers and images rather than semantic similarities, making it much more difficult for the model as the size of the image set increases. It is also observed that AVG demonstrated the best performance across different sizes of image sets, confirming its robustness.

\begin{table}[t]
\renewcommand\arraystretch{1}
  \centering
 \caption{Comparison of the typical methods from three paradigms on performance, applicability to large-scale retrieval, and efficiency.}
 \vspace{-0.5em}
    \scalebox{0.8}{
    \begin{tabular}{ccccccc}
    \toprule
    \multirow{2}*{ Paradigm}&\multicolumn{1}{c}{\multirow{2}*{Method}} &\multicolumn{3}{c}{\makecell[c]{Performance}}&\multirow{2}*{\makecell[c]{Large-scale \\ retrieval}}&\multicolumn{1}{c}{Efficiency} \\ \cline{3-5} \cline{7-7}
         &&R@1&R@5&R@10&&latency\cr
    \toprule
    One-tower&BLIP~\cite{li2022blip}&87.6&97.7&99.0&\textcolor{red}{\ding{55}}&0.00007/s \cr
    Two-tower&CLIP~\cite{radford2021learning}&58.4&81.5&88.1&\textcolor{green}{\ding{51}}&4.32/s \cr
    Generative&AVG&62.8&85.4&91.2&\textcolor{green}{\ding{51}}&16.8/s \cr
    \toprule
    \end{tabular}}
    \vspace{-1em}
    \label{tab:paradigm comparison}
\end{table}

\subsection{Paradigm Comparison (RQ2)}
As a new approach, we also compared AVG with the classical one-tower framework and two-tower framework. We summarize the three different paradigms (BLIP, CLIP, AVG) from the aspects of performance, large-scale retrieval, and efficiency (latency). The performance is evaluated on Flickr30K in terms of R@1, R@5, and R@10. The ``large-scale retrieval'' means the paradigm could be applicable to large-scale image sets or not. Efficiency is measured by the number of queries it can process per second using an image set containing 0.3 million images. The results are summarized in Table~\ref{tab:paradigm comparison}, and we gained the following findings.

\noindent$\bullet\quad$ There is no doubt that the one-tower framework excels at ranking a few numbers of images, and BLIP achieves 99.0 in terms of R@10. The one-tower framework applied a cross-attention mechanism and fine-grained interactions to match the text and image, where the tokens in the text and regions in the image are fully attended. However, as the old saying goes, ``Getting something for nothing''. The fine-grained interaction largely decreases the efficiency, making the one-tower framework inapplicable for large-scale retrieval.

\noindent$\bullet\quad$ Both the two-tower framework and AVG are applicable to large-scale retrieval. For CLIP, it encodes the text and an image into embedding vectors and adopts the dot production to calculate similarities. This simple operation enables CLIP for large-scale retrieval and could process xx textual queries per second with 0.3 million images. Compared to CLIP, our AVG achieves better efficiency, processing xx textual queries per second. This benefits from the generative paradigm, keeping efficiency constant regardless of the size of image sets. More importantly, AVG achieves a better performance in R@1 and R@5. While decoding vokens step by step, AVG could attend tokens and thus introduce a token-level interaction.

\begin{table}[t]
\renewcommand\arraystretch{1}
  \centering
    \caption{ Ablation study on key components of AVG.}
    \vspace{-1em}
    \scalebox{1.0}{
    \begin{tabular}{cccc}
    \toprule
    \multicolumn{1}{c}{\multirow{2}*{Method}}
    &\multicolumn{3}{c}{\makecell[c]{Flickr30K}}\\\cline{2-4}
         &R@1&R@5&R@10\cr
    \toprule
    AVG&62.3&83.5&86.6\cr\toprule
    w/o cross-modal alignment&56.8&76.5&79.2\cr
    w/o voken embedding&62.4&82.7&85.8\cr 
    w/o discriminative loss&60.1&81.9&85.0\cr\toprule
    \end{tabular}}
    \vspace{-1em}
    \label{tab:ablation study}
\end{table}

\subsection{Ablation Study (RQ3)}
In AVG, the cross-modal alignment loss is introduced in the training of the image tokenizer to inject semantics into vokens, and the voken embeddings are transformed to expand the token vocabulary. Do the cross-modal alignment and voken embedding work? In the token-to-voken generative training phase, the discriminative loss is used to modify the solely generative training. We conducted ablation studies by evaluating the following variants: 1) ``w/o cross-modal alignment''. We removed the $\mathcal{L}_{align}$ from the image tokenizer's training. 2) ``w/o voken embedding''. While expanding the language model's vocabulary, we randomly initialized embedding for tokens rather than copying the trained voken embeddings. 3) ``w/o discriminative loss''. We eliminated the loss $\mathcal{L}_{dis}$ and trained the language model only via $\mathcal{L}_{gen}$. By analyzing the results in Table~\ref{tab:ablation study}, we had the following observations.

Removing the cross-modal alignment loss significantly degrades AVG's performance, with R@1 dropping from 62.3 to 62.4. This highlights the importance of aligning high-level semantics in the image tokenizer stage. Interestingly, removing the voken embedding only slightly damages performance, as the voken embeddings can still be well learned in the token-to-voken generate training. Comparing the two variants also provides valuable insight, showing that the cross-modal alignment in the tokenizer stage primarily ensures effective voken ID assignment rather than learning semantic embeddings. Additionally, removing the discriminative loss results in a significant decrease in performance, emphasizing the necessity and effectiveness of modifying the generative training in the retrieval task.

\begin{table}[t]
\renewcommand\arraystretch{1}
  \centering
    \caption{Performance versus the codebook size $N$ on Flickr30K. The best results are marked in Bold.}
    \vspace{-1em}
    \scalebox{1.0}{
    \begin{tabular}{cccc}
    \toprule
    \multicolumn{1}{c}{\multirow{1}*{Codebook Size}}
         &R@1&R@5&R@10\cr
    \toprule
    256&60.8&82.0&84.8\cr
    512&61.1&82.8&86.1\cr 
    1,024&\textbf{62.3}&\textbf{83.5}&\textbf{86.6}\cr
    2,048&61.2&83.0&86.0\cr\toprule
    \end{tabular}}
    \vspace{-1em}
    \label{tab:codebook size}
\end{table}

\subsection{In-depth Analysis}
\subsubsection{\textbf{Analysis on visual codebook}}
Discretizing images into vokens is a complex task that should not be underestimated. It involves representing a large number of images using a limited number of vokens, which is also could be regarded as a compression problem. For instance, in Flickr30K, there are approximately 30,000 images, and the codebook size $M$ needs to be determined. We reported the performance versus the codebook size $M$ in Table~\ref{tab:codebook size}.

By analyzing the results, we gained the following findings. When the codebook size $N$ is set to 256, there is a significant drop in performance. This is likely due to the challenge of representing 30,000 images with only 256 vokens. This is supported by the observation that increasing the codebook size $N$ leads to improved performance, as a larger visual codebook provides more flexibility to represent various images. However, when the codebook size increases from 1,024 to 2,048, the performance decreases. We believe that a larger visual codebook may reduce compression difficulty, allowing the tokenizer to find a shortcut to satisfy tokenization loss, but it may not be suitable for subsequent token-to-voken generation.

\subsubsection{\textbf{Analysis on voken length}}
In section 4.6.1, we primarily focus on determining the optimal visual codebook size for a collection of images. Similarly, we also address the question of how many vokens should be used to represent an image. To investigate this, we presented the performance of the voken length in Table~\ref{tab:voken length}.

\begin{table}[t]
\renewcommand\arraystretch{1}
  \centering
    \caption{Performance versus the voken length $M$ on Flickr30K. The best results are marked in Bold.}
    \vspace{-1em}
    \scalebox{1.0}{
    \begin{tabular}{cccc}
    \toprule
    \multicolumn{1}{c}{\multirow{1}*{Voken Length}}
         &R@1&R@5&R@10\cr
    \toprule
    2&45.7&73.1&80.6\cr
    4&62.3&83.5&86.6\cr 
    6&63.2&84.0&86.7\cr
    8&62.1&82.6&85.2\cr
    10&\textbf{64.1}&\textbf{84.2}&\textbf{87.1}\cr\toprule
    \end{tabular}}
    \vspace{-1em}
    \label{tab:voken length}
\end{table}

\begin{table}[!t]
\renewcommand\arraystretch{1}
  \centering
    \caption{Performance with different autoregressive model backbones on Flickr30K. The best results are marked in Bold. ``Full'' refers to fine-tuning all parameters, while ``LoRA'' refers to partially fine-tuning with the LoRA~\cite{hu2021lora} technique.}
    \vspace{-1em}
    \scalebox{1.0}{
    \begin{tabular}{cccccc}
    \toprule
    \multicolumn{1}{c}{\multirow{1}*{Backbone}}&Params
         &Finetuning&R@1&R@5&R@10\cr
    \toprule
    T5-base&220M&Full&62.3&83.5&86.6\cr
    T5-large&800M&Full&64.0&84.5&87.6\cr 
    T5-XL&3b&Full&\textbf{64.6}&\textbf{84.9}&\textbf{88.0}\cr
    LLaMA-7b&7b&LoRA&54.8&80.8&85.9\cr \toprule
    \end{tabular}}
    \vspace{-1em}
    \label{tab:model size}
\end{table}

The results presented in Table~\ref{tab:voken length} indicate that voken length 2 performs the worst compared to other variants, suggesting that only two vokens are insufficient to effectively represent an image's content. An increase in voken length to 4 results in a significant improvement, with voken length 6 achieving a better performance. This is attributed to the ability of more vokens to better capture the image's content. Although the voken length of 8 results in a slight performance decrease, a voken length of 10 consistently achieves the best performance. However, due to resource constraints, we did not further increase the voken length. It is worth mentioning that as voken length increases, the generation of the entire vokens of an image becomes more difficult. Besides, more vokens lead to slower generation speeds. Therefore, based on these results, the voken lengths 4 and 6 are considered optimal values.

\begin{table}[t]
\renewcommand\arraystretch{1}
  \centering
    \caption{Performance of AVG versus the beam size. The best results are marked in Bold. ``-'' denotes inapplicable results.}
    \vspace{-1em}
    \scalebox{1.0}{
    \begin{tabular}{cccc}
    \toprule
    \multicolumn{1}{c}{\multirow{1}*{Beam Size}}
         &R@1&R@5&R@10\cr
    \toprule
    1&44.2&-&-\cr
    5&60.6&78.1&-\cr 
    10&62.3&83.5&86.6\cr
    20&62.7&85.1&90.2\cr
    30&62.7&\textbf{85.5}&91.1\cr
    40&\textbf{62.8}&85.4&\textbf{91.2}\cr
    50&\textbf{62.8}&85.4&91.1\cr
    \toprule
    \end{tabular}}
    \vspace{-1em}
    \label{tab:beam size}
\end{table}

\begin{figure}[t]
\centering
\includegraphics[width=0.65\linewidth]{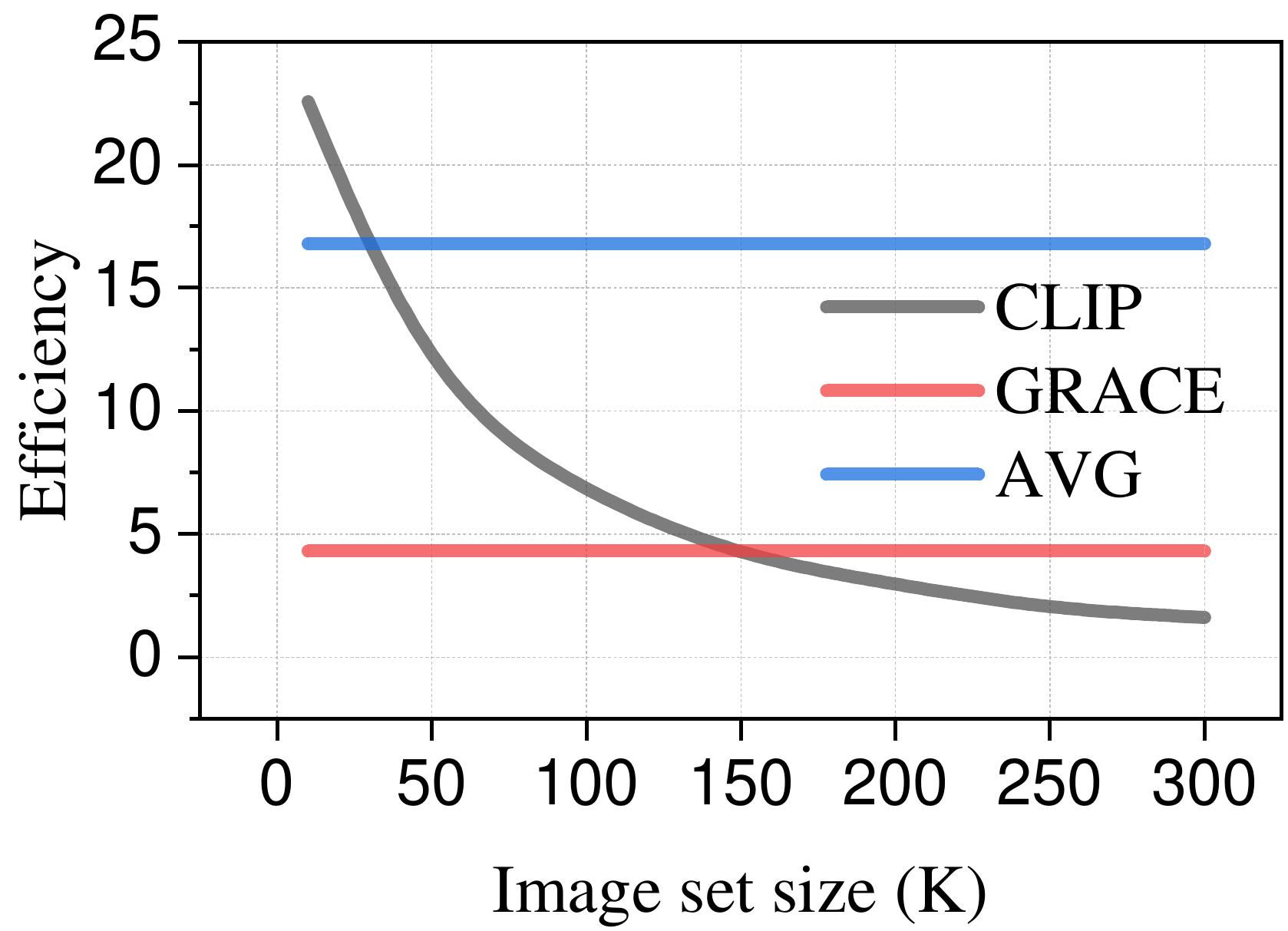}
\vspace{-1em}
\caption{The efficiency of {CLIP} (two-tower), {GRACE} (generative), and AVG (generative) varies with image set size, measured in terms of queries processed per second. {AVG} demonstrates superior efficiency with large image sets.}
\label{efficiency}
\end{figure}

\begin{figure*}[t]
\centering
\includegraphics[width=1.0\linewidth]{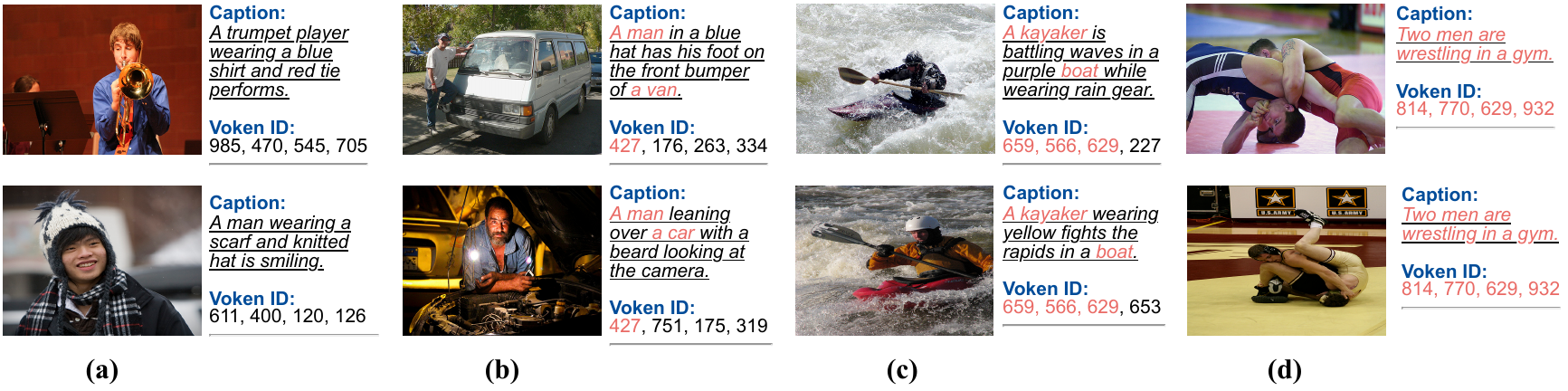}
\vspace{-2em}
\caption{Cases of image tokenization. Four pairs of images to illustrate our cross-modal aligned image tokenizer. In each pair, semantically related words in the caption are highlighted in red. Visual and semantic similar images are assigned similar voken IDs. (a) represents a pair of irrelevant images, while (d) depicts the pair of images with the highest similarity.}
\label{cases}
\end{figure*}

\subsubsection{\textbf{Analysis on autoregressive model size}}
In the token-to-voken training stage, we expanded the vocabulary of generative language models by adding vokens and continued training the language model with text-image pairs. To evaluate the impact of different language models on performance, we conducted experiments using various backbones, such as T5-base, T5-large, T5-XXL, and LLaMA-7b. We utilized the LoRA~\cite{hu2021lora} technique to fine-tune LLaMA due to limited computing resources. The results are summarized in Table~\ref{tab:model size}, and we obtained the following findings.

It has been observed that increasing the model size leads to a certain performance improvement among the T5 series. It is reasonable to expect that a larger language model generally possesses a more powerful semantic understanding ability. In addition to the pretrained knowledge, larger model sizes provide more space for token-to-voken training. An unexpected finding is the significant drop in performance with LLaMA-7b. We have identified two potential reasons for this. Firstly, the most crucial aspect of token-to-voken training is aligning the embedding space between tokens and vokens. Since only a few parameters of the LLaMA-7b could be tuned, this affects the alignment between tokens and vokens. Secondly, it is possible that T5's encoder-decoder architecture is more suitable for generative cross-modal retrieval compared to LLaMA's decoder-only architecture.

\subsubsection{\textbf{Analysis on beam search}}
AVG relies on beam search to generate a ranked list of images, and it is important to investigate how the size of the beam affects the retrieval metrics R@1, 5, and 10. We summarized the performance of AVG at different beam sizes in Table~\ref{tab:beam size}. When using beam search, only the number of sequences smaller than the beam size are returned. Therefore, a beam size of 1 only provides the R@1 result. Our findings show that the greedy search (beam size=1) performs poorly, but performance gradually improves as the beam size increases. This is because a larger beam size reduces the possibility of missing the correct image. The results suggest that a larger beam size is needed to achieve a better R@10, but there is not much difference between beam sizes of 30, 40, or 50. And a larger beam size also reduces efficiency. Therefore, a beam size of 20 or 30 is suitable, considering both effectiveness and efficiency.

\subsubsection{\textbf{Analysis on efficiency}}
As we explore a new approach to cross-modal retrieval, it is crucial to consider efficiency as a key evaluation factor. In our study, we conducted a comparison of AVG with CLIP (two-tower) and GRACE (generative) using different image set sizes. Efficiency was measured by query latency, specifically the number of queries the model could process per second. The results are depicted in Figure~\ref{efficiency}.

As the size of the image set increases, it is observed that the efficiency of CLIP decreases. This is because CLIP needs to rank images by matching the textual query with each image in the set. In contrast, AVG and GRACE maintain consistent efficiency regardless of the image size. This demonstrates a significant advantage of the generative paradigm for cross-modal retrieval. Additionally, within the generative paradigm, AVG shows a notable efficiency advantage over GRACE. This is because GRACE relies on a multimodal language model to memorize images, which tend to have a large size. On the other hand, AVG learns the image tokenizer and only requires the small language model for inference.

\subsubsection{\textbf{Cases of image tokenization}}
We showed cases in Figure~\ref{cases} to demonstrate the image tokenizer in AVG. It is observed that similar images are assigned similar vokens. For instance, the two images in Figure~\ref{cases} (a) are assigned completely different vokens, whereas the two images in Figure~\ref{cases} (b) share the same initial voken. Additionally, we observed that AVG's image tokenizer takes into account the high-level semantics of the image. The two images in Figure~\ref{cases} (c) depict a kayaker in a boat and share similarities, such as the subject matter. However, there are some differences as detailed in captions, such as the color of the boat and clothes, as described in the caption. As a result, the two images are assigned similar vokens ``659, 566, 629, 227'' and ``659, 566, 629, 653'' but not the same. The two images in Figure~\ref{cases} (d), despite having some differences in background, color, and action, are assigned the same vokens. This is because the two images have the same caption, and our image tokenizer is able to focus on the high-level semantic content of the images and filter out irrelevant visual information for retrieval.

\section{CONCLUSION AND FUTURE WORK}
In this paper, we presented a novel autoregressive voken generation method, AVG, designed for the text-to-image retrieval task. Unlike previous discriminative methods that focus on matching text with images, AVG takes a different approach by reformulating text-to-image retrieval as token-to-voken generation. To tackle the unique challenges of cross-modal retrieval, AVG introduces cross-modal alignment image tokenization, which involves discretizing images into vokens while considering both low-level visual information and high-level semantics. Additionally, to address the learning gap in generative training for retrieval targets, AVG incorporates a discriminative loss to adjust the learning direction during the training stage. Our experiments demonstrate that AVG significantly outperforms previous generative paradigms in cross-modal retrieval, thanks to the powerful image tokenizer and modified learning loss of AVG. Furthermore, AVG exhibits superior effectiveness and efficiency, showcasing advantages over both one-tower and two-tower frameworks.

In the future, we plan to further explore this topic from the following perspectives. Firstly, AVG involves two separate stages, image tokenization, and generation training, to facilitate generative cross-modal retrieval. However, these two disconnected stages may result in certain drawbacks. For instance, the assigned vokens may not be optimal for the subsequent generative training. One potential solution is to integrate the two training stages, creating a more seamless end-to-end process. Secondly, image tokenization for generative retrieval entails compressing a collection of images into a sequence of vokens using a visual codebook of a specific size. There may be a strict mathematical relationship among these three variables, which could help determine the optimal parameters for different image sets.
\bibliographystyle{ACM-Reference-Format}
\bibliography{sample-base}

\end{document}